\documentclass{article}
\usepackage{icmc2023template}
\usepackage{times}
\usepackage{ifpdf}
\usepackage{soul}
\usepackage[english]{babel}
\usepackage{xcolor}

\def\papertitle{A Comprehensive Survey for Evaluation Methodologies of AI-Generated Music}
\def\firstauthor{Zeyu Xiong}
\def\secondauthor{Weitao Wang}
\def\thirdauthor{Jing Yu}
\def\fourthauthor{Yue Lin}
\def\fifthauthor{Ziyan Wang}

\newif\ifpdf
\ifx\pdfoutput\relax
\else
   \ifcase\pdfoutput
      \pdffalse
   \else
      \pdftrue
  \fi
\fi

\ifpdf 
  \usepackage[pdftex,
    pdftitle={\papertitle},
    pdfauthor={\firstauthor, \secondauthor, \thirdauthor},
    bookmarksnumbered, 
    pdfstartview=XYZ 
   ]{hyperref}

  \usepackage[pdftex]{graphicx}
  \graphicspath{{./figures/}}
  \DeclareGraphicsExtensions{.pdf,.jpeg,.png}

  \usepackage[figure,table]{hypcap}

\else 
  \usepackage[dvips,
    bookmarksnumbered, 
    pdfstartview=XYZ 
  ]{hyperref}  

  \usepackage[dvips]{epsfig,graphicx}
  \graphicspath{{./figures/}}
  \DeclareGraphicsExtensions{.eps}

  \usepackage[figure,table]{hypcap}
\fi

\hypersetup{
    colorlinks,%
    citecolor=black,%
    filecolor=black,%
    linkcolor=black,%
    urlcolor=black
}

\title{\papertitle}

\fiveauthors
  {0.07in}
  {\firstauthor} { \small Chilli Chicky Band \\ \small Computational Media and Arts Thrust \\ \small \textit{The Hong Kong University of} \\\small \textit{Science and Technology (Guangzhou)} \\ \small Guangzhou, China \\%
    {\tt\small \href{mailto:zxiong666@connect.hkust-gz.edu.cn}{zxiong666@connect.hkust-gz.edu.cn}}}
  {\secondauthor} {\small Chilli Chicky Band \\ \small Earth, Ocean and Atmosphere Thrust \\ \small \textit{The Hong Kong University of} \\\small \textit{Science and Technology (Guangzhou)}  \\ \small Guangzhou, China \\%
    {\tt\small \href{mailto:weitaowangwtw@outlook.com}{weitaowangwtw@outlook.com}}}
  {\thirdauthor} { \small Chilli Chicky Band \\ \small Carbon Neutrality and \\ \small Climate Change Thrust \\ \small \textit{The Hong Kong University of} \\\small \textit{Science and Technology (Guangzhou)}  \\ \small Guangzhou, China \\%
    {\tt\small \href{mailto:jyu336@connect.hkust-gz.edu.cn}{jyu336@connect.hkust-gz.edu.cn}}}
  {\fourthauthor} { \small Chilli Chicky Band \\ \small Computational Media and Arts Thrust \\ \small \textit{The Hong Kong University of} \\\small \textit{Science and Technology (Guangzhou)} \\ \small Guangzhou, China \\%
    {\tt\small \href{mailto:ylin491@connect.hkust-gz.edu.cn}{ylin491@connect.hkust-gz.edu.cn}}}
  {\fifthauthor} { \small Chilli Chicky Band \\ \small Computational Media and Arts Thrust \\ \small \textit{The Hong Kong University of} \\\small \textit{Science and Technology (Guangzhou)} \\ \small Guangzhou, China \\%
    {\tt\small \href{mailto:zwang082@connect.hkust-gz.edu.cn}{zwang082@connect.hkust-gz.edu.cn}}}

\begin{document}
\capstartfalse
\maketitle
\capstarttrue
\begin{abstract}
In recent years, AI-generated music has made significant progress, with several models performing well in multimodal and complex musical genres and scenes. While objective metrics can be used to evaluate generative music, they often lack interpretability for musical evaluation. Therefore, researchers often resort to subjective user studies to assess the quality of the generated works, which can be resource-intensive and less reproducible than objective metrics. This study aims to comprehensively evaluate the subjective, objective, and combined methodologies for assessing AI-generated music, highlighting the advantages and disadvantages of each approach. Ultimately, this study provides a valuable reference for unifying generative AI in the field of music evaluation.
\end{abstract}

\section{Introduction}\label{sec:introduction}
With the development of artificial intelligence generation technology, a large amount of work and applications have been generated for intelligent music generation~\cite{kaliakatsos2020artificial, ji2020comprehensive, cao2023comprehensive, dai2021controllable}. In particular, Music generation can be further divided into two types: the symbolic domain and the audio domain. Music generation in the symbolic domain is stored in MIDI format, and its textual and sequential data nature facilitates its applications (e.g., MidiNet~\cite{yang2017midinet}, MuseGAN~\cite{dong2018musegan}, BandNet~\cite{zhou2018bandnet} and TeleMelody~\cite{ju2021telemelody}) in major deep learning models (e.g., LSTM~\cite{eck2002first, wu2019hierarchical}, autoencoder~\cite{bretan2016unit}, RBM~\cite{hinton2006reducing}, and GAN~\cite{goodfellow2020generative}). For the audio domain, it is also possible for the analysis of the different bands according to the characteristics of the audio to obtain vectorized data for model training (e.g., Jukebox~\cite{dhariwal2020jukebox}, WaveNet~\cite{oord2016wavenet}). In addition to generating music from MIDI datasets or audio datasets, many works have started to look for connections between multimedia. For example,  MusicLM~\cite{agostinelli2023musiclm} generates music from text, and BGT-G2G~\cite{xiong2022retaining} generates music from images. All the above-mentioned works reach a certain level of accepted musicality. However, these ratings are either entirely referenced to parameters such as the accuracy of model training or are subjective ratings that rely entirely on user study. 

\begin{figure}[h]
\centering
\includegraphics[width=0.9\columnwidth]{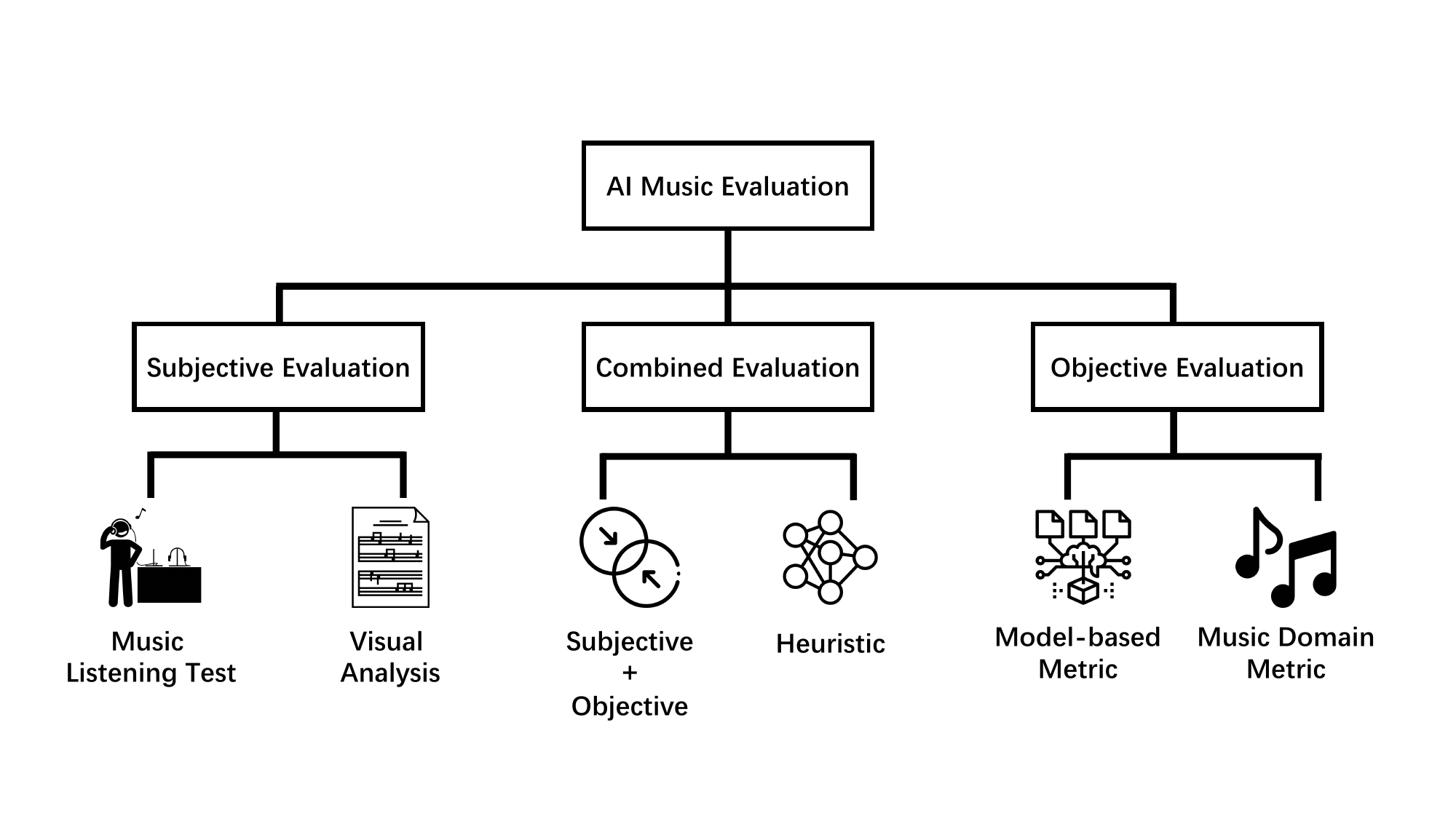}
\caption{Overview Structure of The Survey: AI Music Evaluation Methods\label{fig:structure}}
\end{figure}

Due to the different experimental processes and judgment criteria for subjective ratings, the objective model training metrics do not directly represent subjective feelings. There has not been a broad consensus on the evaluation of such generative models for a long time, resulting in a great challenge for music generation models in determining evaluation criteria~\cite{borji2019pros, jordanous2012standardised}. While subjective evaluation is usually better suited for evaluating generative models, it can be resource-intensive, and there are no uniform criteria. In contrast, objective methods, even if easy to implement, are usually less explanatory. To this end, we are dedicated to performing a survey for evaluation methodologies of AI-generated music and providing reference values for designing more scientific and effective evaluation methods in the future. Figure~\ref{fig:structure} shows the overview structure of this survey. We separate our survey into three main categories: (1) subjective evaluation, (2) objective evaluation, and (3) combined evaluation.

Our contributions to this work are listed below:

\begin{enumerate}
    \item We provide a classification reference scheme of evaluation for creators in the field of AI music.
    \item We provide a reference value for the unification of generative AI in the field of music evaluation.
\end{enumerate}

\section{Related Work}\label{sec:related-work}

In recent years, artificial intelligence (AI) systems have been increasingly involved in various applications of music composition, ranging from entertainment to therapeutic uses. However, as the popularity of AI-generated music grows, an increasing number of issues have come to light (e.g., concerns about plagiarism in AI-generated music content \cite{yin2021good}, etc.). Therefore, to ensure the success of these applications, it is crucial to establish effective evaluation metrics. 

Existing surveys on the evaluation method of AI-generated music predominantly categorize the evaluation methodologies into two primary classifications: subjective evaluations and objective evaluations. 
Subjective evaluation methods entail the solicitation of human listeners to provide ratings based on specific criteria, such as musicality, novelty, or emotional impact. Scholars such as Ji et al. \cite{ji2020comprehensive} and Zhao et al. \cite{zhao2022review} have notably emphasized the use of listening tests as a widely employed approach to assess the melodic output of AI-generated music. Given the inherent subjectivity inherent in music appreciation, strict standardization of such evaluations remains a challenging endeavor, as noted by Yamshchikov et al. \cite{yamshchikov2020music}.

On the other hand, objective evaluation methods center on quantitative measurements applied to both the generated music and the underlying generative models. Ji et al. \cite{ji2020comprehensive} elaborate on the objective evaluation as the quantitative measurement for both the generated music and the generative model. Notably, Theis et al. \cite{theis2015note} conducted evaluations focusing on generative model performance from the perspective of log-likelihood, while Civit et al. \cite{civit2022systematic} extended this evaluation to encompass considerations of dataset, code integration, and system structure.  
The selection of subjective or objective assessment methods often hinges on the specific purposes and criteria of measurement. Objective evaluation methods are commonly utilized in tasks such as classification, prediction, and recommendation, whereas subjective evaluations are more frequently employed to assess the quality of the generated musical content. In the following subsections, we delve into existing subjective and objective evaluation methods, and we also explore the amalgamation of these approaches. Furthermore, we analyze the application scenarios wherein each evaluation methodology finds relevance and effectiveness in the context of AI-generated music assessment.

\section{Subjective Evaluation}

Subjective evaluation of generated music is predominantly reliant on assessments provided by human listeners emphasizing their satisfaction. Due to the subjective evaluative nature of music in the real world, even though it tends to be more resource-intensive and less reproducible than objective evaluation, subjective evaluation is still an integral part of the process of AI music generation evaluation. Among the existing generative models, music listening tests and visual analysis are the two most important parts.

\subsection{Music Listening Test}

The music listening test is the most common method in subjective evaluation. Such evaluations are commonly conducted through two approaches: the musical Turing test or subjective query metrics based on modeled compositional theory~\cite{yang2020evaluation}. Generally, a validated music listening test should require these conditions~\cite{ribeiro2011crowdmos}: (1)  The experiment was conducted in a controlled environment with specific acoustic characteristics and equipment, (2) The music knowledge level of subjects was evenly distributed, including both music amateurs who are lacking in music knowledge and experts in the field of music composition, (3)  Each subject received the same instructions, and (4) Each subject received statistically significant results.


\subsubsection{Musical Turing Test}\label{sec: Musical Turing Test}

Musical Turing Test measures the extent to which the generated music is indistinguishable from human-composed music. For example, Nadeem et al.~\cite{nadeem2019let} tested 28 users from the aspects of accuracy and preference to evaluate the generator output of a deep learning architecture combined with the proposed musical data. Although 57\% of them correctly recognized that the music piece was generated by the computer, some claimed that the decision process was difficult. Ferreira et al.~\cite{ferreira2023generating} invited 117 participants to differentiate the music composed by humans or five models (three transformers and two RNN models). What this study did good was that it divided all members into three groups according to their experience of classical music. It proved that people with better musical sensitivity had a higher correct rate of distinguishment. 

To demonstrate statistically significant differences in evaluation results, hypothesis testing is usually performed (e.g., t-test~\cite{donahue2019lakhnes}, h-test~\cite{chi2020generating}, etc.).

\subsubsection{Subjective Query Metrics}

Nadeem et al.~\cite{nadeem2019let} and Ferreira et al.~\cite{ferreira2023generating} also ask if the audiences love experimental music by a binary question. However, the answers cannot quantify the degree of their preferences with less representative. Therefore, more metrics should be considered during the evaluation of different models, which need to be specifically explained. In the study conducted by Chu et al.~\cite{chu2022empirical}, 100 people participated in evaluating their interests in the music performed by different transformer models. This survey ruled out nine parameters (including overall creativity, naturalness, melodiousness, richness, rhythmicity, correctness, structures, and coherence) in a 7-point Likert scale. Take creativity as an example. It was described as the degree of novelty, value, and origin of the music pieces. In this way, participants can evaluate the music more specifically after a quantification process. Compared with the study, Hernandez-Olivan et al.~\cite{hernandez2022subjective} considered the melody, harmony, and rhythm of the music on a 5-point Likert scale. But they designed respective questions for different groups. Specifically, for people with less music knowledge, three parameters can be reflected by two questions. As for professionals with a solid foundation in music theory, they should answer six questions to determine those metrics. What’s more, the experience of users should be paid more attention. Like the duration and the number of music pieces, these factors possibly make people fatigued, which may lead to a higher deviation in the results.

\subsection{Visual Analysis}

For visual analysis, the involvement of a music expert is often required. It is up to the expert to analyze the score, chord progression sheet, piano roll, etc., after visualization. For example, Dong et al.~\cite{dong2018musegan} analyzed the stability, fluidity, and musicality of melodies generated in chordal and rhythmic patterns in music. The waveform and spectrogram of the audio samples are also considered indicators for subjective evaluation. For example, Engel et al.~\cite{engel2017neural} show each note as a ``Rainbowgram'', which is a visualization technique to show the relationship between time and frequency.

\subsection{Summary}

In conclusion, we believe that a good subjective evaluation should not only contain a precise design but also consider the users’ characteristics. In this way, target metrics can be better determined while participants react in a comfortable environment, which can also contribute to the development of improved algorithms and models in the field.

Besides, although the subjective judgment is indispensable because of the artistic subjectivity of music, the resources expended behind it are enormous. At the same time, it is difficult to ensure the reproducibility as well as the stability of the experiments. Therefore, with the help of some objective quantitative indicators, it would be helpful to analyze the quality of music generation in a more scientific way.

\section{Objective Evaluation}

The objective evaluation involves using computational techniques to analyze the music and generate objective measures of its quality. Dong et al.~\cite{dong2018musegan} and Sturm~\cite{sturm2017taking} have used evaluation metrics based on probabilistic measures such as likelihood and density estimates (especially in the field of image generation~\cite{theis2015note}), yet whether there is a direct link between good or bad models and music quality is not yet known. Besides, metrics such as model metrics and music metrics are often used. For example, researchers may use metrics such as pitch entropy, chord progression complexity, or rhythmic variance to evaluate the music quality. We discuss the application of these metrics in detail in this section.


\subsection{Model-based Metrics}

Model-based Metrics refer to the general generative model evaluation metrics that do not contain music domain knowledge. Some common model-based metrics include \textit{training loss}, \textit{precision}, \textit{recall}, \textit{f1 score}, etc. Other metrics like the \textit{chord prediction accuracy}~\cite{lim2017chord}, \textit{style likelihood}~\cite{brunner2018midi}, and \textit{reconstruction accuracy}~\cite{roberts2018hierarchical} are also applied for the objective evaluation.

Model-based evaluation methods are also limited to specific models or methods without strong universality because the methods and models of different generation systems are very different. Bretan et al.~\cite{bretan2016unit} considered a unit to be a variable length number of measures of music, and utilized objective metrics to assess the generative model, such as mean rank and accuracy, by evaluating the rank of the target unit. This is a specific evaluation metric based on model characteristics, not general metrics. Thus, a model-based metric inspired by domain knowledge is not universal but performs well on a particular task, even though its interpretability remains questionable in terms of music quality.

\subsection{Music Domain Metrics}

Music Domain Metrics (MDM) refers to the evaluation index under the domain knowledge of music, such as volume, pitch, chord, score, etc. Ji et al.~\cite{ji2020comprehensive} categorized these metrics into 4 categories: (1) pitch-related, (2) rhythm-related, (3) harmony-related, and (4) style-related. 

\subsubsection{Pitch \& Rhythm Related Metrics}

Widely-used pitch-related and rhythm-related metrics include scale consistency, tone spam, consecutive pitch repetitions, qualified rhythm frequency, rhythm variations, etc~\cite{mogren2016c, trieu2018jazzgan}. For the state-of-the-art metric design, Yang and Alexander~\cite{yang2020evaluation} propose a set of musicological objective assessment metrics, using which the output of the music generation model can be evaluated and compared. These metrics were validated in experiments and are reproducible. The proposed features include pitch counts, pitch category histograms, pitch shift matrices, pitch spans, average pitch intervals, note counts, average repetition intervals, note length histograms, and note length shift matrices. 

\subsubsection{Harmony Related Metrics}

Harmony-related metrics focus on measuring harmonic consistency, chord histogram entropy, chord coverage, polyphony (how often two tones are played simultaneously), tone span, etc~\cite{yeh2021automatic}. For example, C-RNN-GAN~\cite{mogren2016c} and  JazzGAN~\cite{trieu2018jazzgan} used harmony-related metrics to measure the compatibility and musicality of generated outputs.

\subsubsection{Style Related Metrics}
In terms of style transfer, ``Style Fit'' (how well the generated music fits the desired style) and ``Content Preservation'' (how much content it retains from the original) are most commonly mentioned~\cite{brunner2018symbolic}. Cifka et al.~\cite{cifka2020groove2groove} proposed a new set of objective evaluation metrics to be used alongside existing metrics. To capture the consistency in the harmonic structure, they preserve the content by calculating the frame-by-frame cosine similarity between chromatic features. For style fit, they collected so-called style profiles~\cite{mckay2004automatic} to measure how well they are matched by the style transfer outputs.



\subsection{Summary}
In this section, we introduce model-based metrics and music domain metrics, as well as state-of-the-art innovative metrics. While the purpose of the above approaches is to reduce the workload of crowd-sourcing through scientific data, the interpretability of the above methods remains to be verified as the quantitative indicators do not fully represent subjective human perceptions. Therefore, studying the interpretability of objective evaluation indicators remains an issue worth exploring~\cite{castelvecchi2016can}.

\section{Combined Evaluation}\label{sec:evaluation}

Combining subjective and objective evaluation methods can be an effective approach to evaluating AI-generated music. Recently, heuristic algorithmic frameworks have also emerged as important tools for combined evaluation. In this section, we discuss how the combined evaluation is performed and tested.

\subsection{Subjective + Objective Evaluation}

Subjective plus objective evaluation refers to evaluation methods that combine subjective user study and objective metrics, and a comparison between the subjective and objective evaluation forms the final conclusion. For example, Zhao et al.~\cite{zhao2019emotional} evaluated the AI-generated music by combining objective musical metrics (polyphony, scale consistency, 3-tone repetitions, and tone span) analysis and subjective query metrics (harmonious, rhythmic, musically structured, and coherent) user study together. Huang et al.~\cite{huang2021modeling} trained a music mashup model and evaluated the outputs by combining objective evaluation (model-based metric plus music domain metric) and subjective listening tests (analyze mean of scores). The combined-style assessment of the above work intercepts the respective strengths of subjective and objective assessments and provides a more comprehensive assessment of the model's strengths and weaknesses in a broader dimension. However, the level of diversity in the database used above poses a challenge to the generalizability of the model. Besides, since the combination of the two also requires aligning final results, there is still no uniformity in the interpretable migration of the objective assessment compared to the subjective assessment.

\subsection{Heuristic Evaluation Framework}
Dervakos et al.~\cite{dervakos2021heuristics} propose a heuristic framework to calculate the frequency of different features by using tools such as the "five-degree circle" to output quantitative scores for each metric. In this framework, the authors define four heuristic objective assessment attributes based on intuition and empirical observations as musicality. However, due to interpretability limitations, the authors still made a subjective assessment to prove the existence of the objective property was meaningful. In the subjective test, over 1,000 users participated in scoring three dimensions: 1) how much they liked the music, 2) how interesting the music was, and 3) the Turing test: whether the composer of the music was a human or a computer. The authors compared the final results of the user survey with the results of their proposed heuristic and eventually found a high degree of similarity in the results between the two, thus demonstrating the significance of the heuristic.

\subsection{Summary}

Subjective + objective evaluation is designed for learning the broad domain of the generated music. Through this method, many work evaluations become more robust. However, we have not yet found a unified assessment paradigm because of the different objectives of the different efforts. The heuristic evaluation framework seems to contribute significantly to mitigating the resource consumption of purely subjective evaluations. While subjective evaluation can provide valuable insights into the general public's perception of creative AI and the evaluation of music, heuristics can be used to evaluate specific features of AI-generated music without the need for comparison between generated and real data. However, the robustness of the method still needs to be compared in parallel with a purely subjective evaluation. This, in turn, gets caught in the trade-off between interpretability and experimental reproducibility.

\section{Discussion}\label{sec:discussion}

As AI music generation continues to evolve, the methods of evaluating these generated outputs must also adapt to the increasing complexity and creativity of the models. There are several challenges and future directions that we have identified in this field.

\subsection{Establishing Standards}
At present, the evaluation of AI-generated music lacks standardization, which results in a process that is inconsistent and lacks a common reference point for stakeholders, including developers, musicologists, and audiences. The creation of a comprehensive, standardized evaluation system would streamline this process, benefitting all parties involved.

The envisioned system would include a set of standard metrics, incorporating both subjective and objective elements, applicable across various AI models and across different music genres. This is vital as different genres of music possess unique characteristics, influencing the type of metrics required for their evaluation.

For instance, the complexity of harmonic structure forms a critical component of classical music evaluation. In contrast, the "catchiness" or melodic hooks are often a major focus in the evaluation of pop music. This diversity necessitates the challenge of formulating genre-specific evaluation metrics, allowing for a fair and accurate appraisal of AI-generated music within the context of its intended genre.

In essence, the development of a standardized evaluation system that accounts for both general musical elements and genre-specific characteristics is a pressing need in the field of AI-generated music.

\subsection{Bridging the Gap between Subjective and Objective Evaluation}
As discussed in Section~\ref{sec:evaluation}, one of the main challenges lies in bridging the gap between subjective and objective evaluations. While subjective evaluation considers the listeners' personal preferences and emotions, objective evaluation relies on mathematical and computational analysis. The challenge is to find a balance and a correlation between these two methods. Future research could focus on developing methods that can effectively combine these two approaches to provide a comprehensive evaluation.

\subsection{Interpretability of Objective Metrics}
Although objective metrics provide a quantitative measure of the quality of AI-generated music, their interpretability remains an issue. Many of these metrics are based on abstract mathematical concepts that may not necessarily correlate with human perception of music quality. Therefore, it is crucial to develop objective metrics that can accurately represent subjective human perceptions and can be easily interpreted in terms of music quality.

\subsection{Evaluating Creativity}
Evaluating creativity in AI-generated music is a complex task as it involves assessing different criteria in different contexts. Some of these criteria are novelty, originality, and value. Novelty refers to the newness or uniqueness of a musical piece. A composition that sounds distinctly different from existing pieces can be considered novel. However, it's essential to remember that novelty alone does not equate to creativity. For instance, random notes played together might be novel but not necessarily creative or enjoyable. Another criterion is originality: it is closely related to novelty, but it adds an extra layer of refinement. An original piece of music introduces something new while also demonstrating an understanding of existing musical traditions and structures. It should show a level of sophistication and skill, breaking from the norm in a purposeful and artful way. Value is the third critical component of creativity. A creative piece of music should be novel and original with value, which can be emotional, cultural, aesthetic, or intellectual. It might be a piece that resonates deeply with listeners, offers a new perspective, or pushes boundaries in the music world. Current evaluation methods may not fully capture these aspects, and different audiences' definition of creativity varies. Therefore, developing methods to evaluate creativity effectively is a significant challenge. This could involve devising new metrics or modifying existing ones to measure these aspects.




\section{Conclusions}
AI music generation is a promising field with significant potential for both creative and technological advancements. The evaluation of AI-generated music, however, is still a challenging and complex task that requires both subjective and objective methods. In this paper, we discussed various evaluation methods, including subjective evaluations like music listening tests and visual analysis, objective evaluations like model-based metrics and music domain metrics, and combined methods. We conducted a comprehensive survey for the evaluation methodologies of AI-generated music; we separated these methods into three parts: (1) subjective evaluation, (2) objective evaluation, and (3) combined evaluation. We discussed in detail the advantages and disadvantages of various evaluation methods and provide a future perspective on the evaluation of generative AI in the music domain. This work also provided insights for the future release of a unified style of assessment method. We also outlined several future directions and challenges in this field, such as establishing standards, bridging the gap between subjective and objective evaluations, and evaluating creativity and different music genres. We believe that addressing these challenges will lead to more reliable and comprehensive evaluation methods for AI-generated music, contributing to the further development of this field.



\bibliography{icmc2023template}

\end{document}